%% file: main.tex
\documentclass[]{aa}

\usepackage[varg]{txfonts}

\usepackage{natbib}
\bibpunct{(}{)}{;}{a}{}{,} % to follow the A&A style

\usepackage{xcolor}

\usepackage[normalem]{ulem}     %Provides the following commands:
\makeatletter
\renewcommand*\aa@pageof{, page \thepage{} of \pageref*{LastPage}}
\makeatother

\include{mod_AR_definitions}

\begin{document}

\title{Bayesian approach for modeling solar active region global magnetic parameters}%

\titlerunning{Modeling solar active region global magnetic parameters}
\authorrunning{Poisson et al.} 

\author{M.~Poisson \inst{\ref{inst1}}
\and F.~Grings \inst{\ref{inst1}}
\and C.H.~Mandrini\inst{\ref{inst1}} 
\and M. L\'{o}pez-Fuentes\inst{\ref{inst1}}
\and P. D\'emoulin\inst{\ref{inst2}, \ref{inst3}}
}

\institute{Instituto de Astronomía y Física del espacio (CONICET-UBA), Buenos Aires and CC. 67, Suc. 28, 1428, Argentina. \email{mpoisson@iafe.uba.ar}\label{inst1} \and LESIA, Observatoire de Paris, Universit\'e PSL, CNRS, Sorbonne Universit\'e, Univ. Paris Diderot, Sorbonne Paris Cit\'e, 5 place Jules Janssen, 92195 Meudon, France. \label{inst2}
\and
Laboratoire Cogitamus, rue Descartes, 75005 Paris, France. \label{inst3} 
}

\date{Received date /
Accepted date }

\abstract 
   % context
{Active regions (ARs) appear in the solar atmosphere as a consequence of the emergence of magnetic flux tubes. The presence of elongated magnetic polarities in line-of-sight (LOS) magnetograms indicates the existence of twist in the flux tubes forming them. These polarity elongations, called magnetic tongues, bias the measurement of AR characteristics obtained during their emergence phase (e.g. their tilt angle and magnetic flux, among others).
In particular, obtaining a good estimation of the tilt angle evolution plays a key role in constraining flux-transport dynamo models.} 
   % Aims
{In this work we aim to estimate the intrinsic properties of the twisted flux tubes, or flux ropes, that form ARs by quantitatively comparing observed LOS magnetograms with synthetic ones derived from a toroidal magnetic flux tube model.} 
   % Methods
{For this reason, we develop a Bayesian inference method to obtain the statistical distributions of the inferred model parameters. 
As an example, we apply the method to NOAA AR 10268. Next, we test the results using a synthetic-AR generator to quantify the effect of small scale perturbations over the inferred parameter distributions.} 
   % Results
{We conclude that this method can significantly remove the effects of magnetic tongues on the derived AR global characteristics, providing a better knowledge of
the intrinsic properties of the emerging flux rope.}  
   % Conclusions
{These results provide a framework for future {analysis of the physical properties of emerging ARs using Bayesian statistics.}}


\maketitle

%%%%%%%%%%%%%%%%%%%%%%%%%%%%%%%%%%%%%%%%%%%%%%%%%%%%%%%%%%

%%%%%%%%%%%%%%%%%%%%%%%%%%%%%%%%%%%%%%%%%%%%%%%%%%%%%%%%%%
\section{Introduction}
\label{sect_Introduction}

%{\S\bf --- ARs as the emergence of magnetic flux-ropes} \\
Active regions (ARs) appear in the solar photosphere as concentrated and compact magnetic field distributions. Their origin can be explained by a dynamo mechanism acting at the bottom of the convective zone (CZ), where the magnetic flux is amplified and its structure is distorted \citep{Fan09}. 
Magnetohydrodynamic (MHD) numerical simulations show that a buoyant instability process can create coherent magnetic flux tubes that rise from the deep layers of the CZ out to the solar atmosphere \citep{Weber13}.  These simulations proved that the magnetic flux tubes should be twisted, forming flux ropes (FRs), in order to maintain {their} cohesion against the plasma vortexes that develop in their wakes \citep{Emonet98,Abbett01,Martinez15}.

%{\S\bf --- Magnetic tongues} \\
In general, the photospheric line-of-sight (LOS) magnetic field of ARs displays a bipolar configuration.  Several asymmetries are observed in the AR polarities as a consequence of the emergence process and the twist of the FRs. The so-called magnetic tongues observed in LOS magnetograms are an evidence of the twist of the emerging FRs. They are characterized by the elongation of the magnetic polarities and the inclination of the polarity inversion line (PIL) with respect to the perpendicular direction to the bipole axis (an axis pointing from the following to the leading polarity
centers). 
{The magnetic flux distribution observed in LOS magnetograms is a combination of both the axial and the azimuthal field components projected in the LOS direction.  During the emergence of the FR, the axial field component is associated to the formation of two main magnetic polarities which separate as the AR evolves. The azimuthal component, which is orthogonal to the axial component, introduces an asymmetric extension of the magnetic polarities which receive the name of magnetic tongues.  They are longer and they have a larger magnetic flux as the emerging flux tube is more twisted} \citep{Lopez-Fuentes00,Luoni11}. 
% Where the azimuthal field is associated to the twist of the field lines around the FR axis  \citep{Lopez-Fuentes00,Luoni11}.
%These tongues are due to the projection of the azimuthal magnetic field of the FR on the LOS direction 
This projection effect is stronger during the emergence phase of the AR and has been reported in many different works \citep[see, e.g., ][]{Luoni11,Mandrini14,Valori15,Yardley16,Vemareddy17,Dacie18,Lopez-Fuentes18}.  Tongues are also present in MHD simulations of FR emergence \citep{Archontis10,Cheung10,MacTaggart11,Jouve13,Rempel14,Takasao15}. 

%{\S\bf --- Tilt angle} \\
Another important property of the emerging FRs is their tilt angle, defined as the inclination of the bipole axis  with respect to the east-west direction. Observations show a tendency of the leading polarity of ARs to be located closer to the solar equator relative to the following polarity, {the so-called Joy's Law \citep{Hale19}}. 
The latitudinal distribution of the tilt {\citep[see the review by][and references therein]{vanDriel15}} plays a central role in flux-transport dynamo models as a fundamental ingredient for the formation and evolution of the polar field \citep[see the review by][and references therein]{Wang17}.

%{\S\bf --- Tilt estimations}\\
There has been an important effort over the past few years to characterize the tilt of ARs with a variety of results \citep{Li12,McClintock13,McClintock14,Wang15,Tlatova18}. 
Tilt angles can be derived with different methods depending on the available observations. The longest databases correspond to white-light (WL) photographic observations \citep[see e.g.,][]{Howard84,Sivaraman93,Baranyi16}.
The WL tilt can be obtained by grouping sunspot umbrae (penumbrae) as leading or following polarities, according to their east-west position, and then computing the area-weighted center of each group. 
Due to the fact that sunspots and pores are present in the strongest magnetic fields, strong magnetic tongues are also expected to modify WL images \citep{Poisson2020b}.

%{\S\bf --- coffe}\\
The use of LOS magnetograms provides a more {precise} separation between leading and following polarities than WL data. The mean location of each main magnetic polarity can be obtained by computing the flux-weighted center of each polarity distribution (i.e., the magnetic barycenters). 
%The presence of magnetic tongues naturally modifies the photospheric magnetic distribution of flux concentrations and, therefore, tilt-angle measurements done directly on LOS magnetograms. 
{During the AR emergence, the elongation of the magnetic polarities produces an asymmetry of the flux concentrations on each polarity that shifts the location of the magnetic barycenters towards the PIL, modifying therefore the tilt-angle measurements done directly on LOS magnetograms.
In other words, the azimuthal field component, responsible of %present on 
the magnetic tongues, systematically affect the estimation of the tilt angle in the early stages of the AR evolution.}
In \citet{Poisson20a} we developed a method, called Core Field Fit Estimator (CoFFE) that succeeds to remove most of the effect of magnetic tongues on the computation of the location of the magnetic barycenters and, hence, allows to obtain an AR tilt-angle that better represents the FR intrinsic tilt. 
 
%{\S\bf --- Aims} \\
In this work we aim to obtain AR global magnetic parameters by modeling LOS magnetograms of an emerging AR with synthetic magnetograms generated from a half-torus  model. 
The inferred parameters can provide a better estimation of the FR intrinsic properties, reducing the effect of the magnetic tongues, since the elongation of the polarities produced by the presence of twist is inherent to our model.

%{\S\bf --- roadmap of this paper} \\
In \sect{data} we describe the LOS magnetograms used, the AR selection criteria, and the AR example selected.  \sect{torus} summarizes the half-torus model and describes how the synthetic magnetograms are built. 
In \sect{modeling} we first introduce the Bayesian probabilistic method and describe the tools used to sample the model parameter space. {Next,}  we apply this method to infer model parameters from AR 10268 observed magnetograms. 
\sect{modelgen} presents an AR simulator used  to explore the limitations of the proposed method. 
Finally, we discuss and conclude on the method implications and its applicability to solar data in \sect{Conclusion}.

%--------------------------------------------------------------------

%%%%FIGURE AR 10268 %%%%%%
\begin{figure}[!t]
\centering
\includegraphics[width=0.45\textwidth]{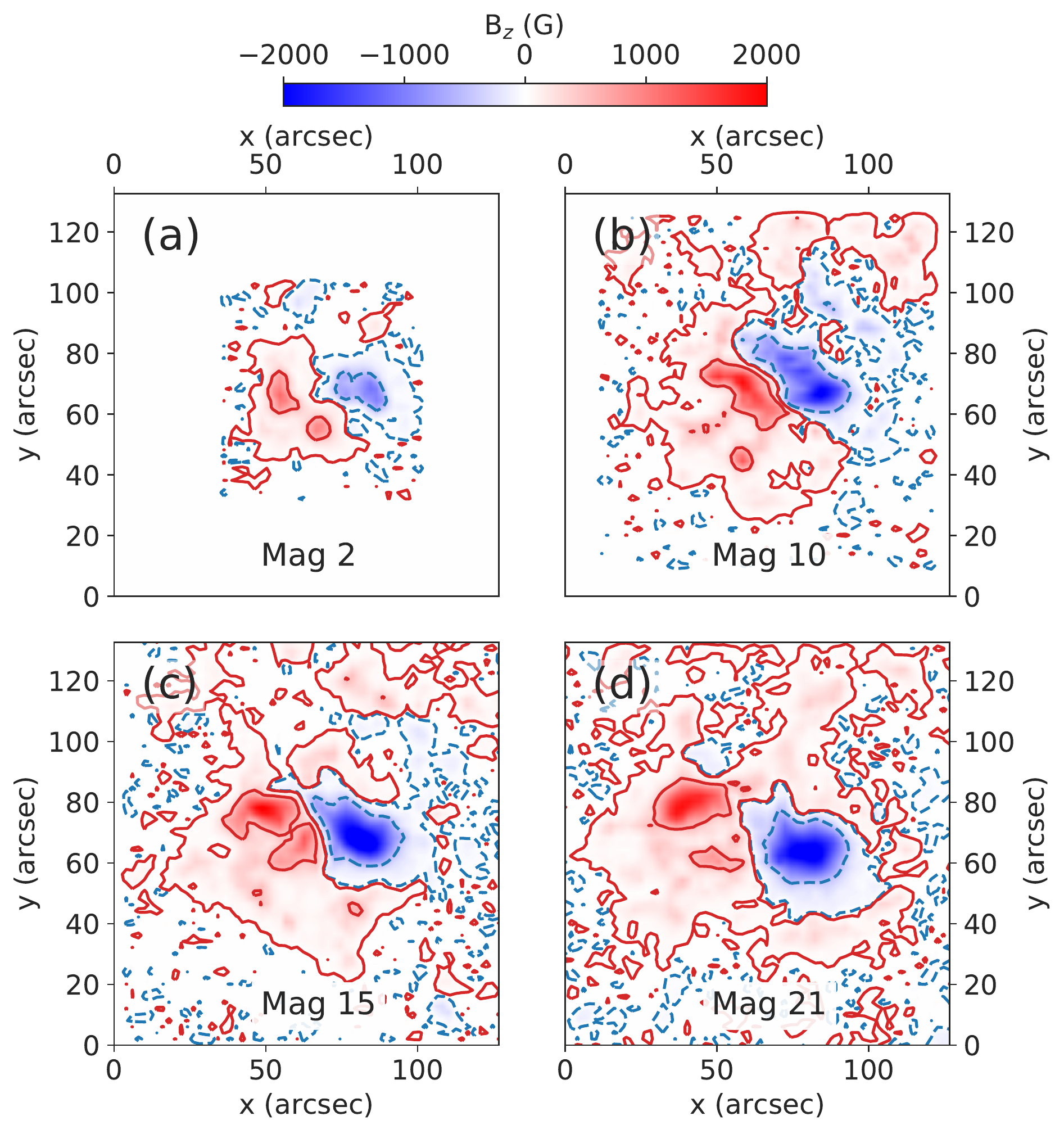}
\caption{LOS magnetograms corresponding to AR 10268 at four different stages of its emergence. The red- and blue-shaded areas represent the positive and the negative LOS magnetic field component. The red (blue) contours correspond to positive (negative) magnetic field with a strength of $20$ G ($-20$ G) and $500$ G ($-500$ G). The magnetograms are limited to square regions adjusted to the evolution of the AR extension. 
}
\label{fig_AR10268}
\end{figure}
%%%%FIGURE AR 10268 %%%%%%

%%%%%%%%%%%%%%%%%%%%%%%%%%%%%%%%%%%%%%%%%%%%%%%%%%%%%%%%%%
\section{Data description and processing}
\label{sect_data}

%{\S\bf --- SOHO/MDI WL and LOS magnetograms} \\
We use full-disk LOS magnetograms obtained with the Michelson Doppler Imager (MDI) on board the Solar and Heliospheric Observatory (SOHO). 
These magnetograms are constructed on board SOHO by measuring the Zeeman effect in right and left circularly polarized light.
The magnetograms from the 96-minute series, obtained from 5-minute averaged magnetograms, have a lower noise level than the 1-hour series and an error per pixel of $\approx 9$ G \citep{Liu04}.
These magnetograms have a spatial resolution of $1.98''$ and are obtained using a CCD of $1024\times1024$ pixels. 

%{\S\bf --- ARs selection and processing LOS magnetograms} \\
From the full-disk magnetograms we select the area containing NOAA AR 10268, covering its full observed emergence corresponding to 66 magnetograms along 5 days of observations. 
We limit the latitudinal and longitudinal range of the selected AR within $-35\degree$ to $35\degree$ from disk center to reduce the projection effect of the LOS magnetic field component { when the observations are close to the solar limb.}
{Panel (a) of the movie mov-fig3.mp4 in the supplementary material shows the evolution of the AR with a cadence of 192 minutes between January 22 and 26, 2003. }

%  {\S}{\bf --- about AR 10268} \\ 
AR 10268  presents a long persistent emergence and it maintains its bipolar configuration along its full evolution.  
{Along the analyzed time span the AR is classified mainly as a $\beta-$type AR. On January 25 a different classification is provided by the NOAA/USAF Active Region Summary, { indicating a $\beta \delta-$type AR; this does not agree with the observed magnetograms because AR 10268 maintains} its simple bipolar configuration. A direct comparison between magnetograms and MDI white light images shows that all the  regions where umbrae share the same penumbra have the same polarity sign, therefore they not correspond to a $\delta-$type classification}.

Despite the asymmetry between the leading and following polarities, the tongues are clearly identified with an elongated pattern associated with the emergence of a negative twisted FR (\fig{AR10268}). 
We also see a sustained rotation of the bipole, suggesting a consistent variation of the intrinsic tilt angle or/and the effect of the continued contraction of the magnetic tongues along the emergence {(see movie mov-fig3.mp4)}.

%{\S\bf --- All data processing} \\
We process the AR magnetograms to construct a data cube using a time step of 192 minutes (that is, we use one third of the available magnetograms).
We decrease the number of magnetograms to reduce the computational time required for each method. This is admissible, since the data cadence has a weak/negligible effect on the results of the method described in \sect{modeling}.
Using standard SolarSoftWare (SSW) tools, we transform the LOS component of the magnetic field to the solar radial direction. 
Since we limit the longitudinal and latitudinal span of the observations, the latter approximation produces no significant effect on the resulting magnetic flux density \citep{Green03}.
Finally, we rotate the set of magnetograms to the time when the AR was located at the central meridian. 
This procedure corrects for the solar differential rotation using the coefficients derived by \citet{Howard90}. 
As a final step, we select a sub-region which encompasses the AR reducing as much as possible the surrounding magnetic field.

%%%%%%%%%%%%%%%%%%%%%%%%%%%%%%%%%%%%%%%%%%%%%%%%%%%%%%%%%%
\section{Half-torus analytical model} \label{sect_torus}

%{\S\bf --- Generalities}\\ 
The FR model, developed by \citet{Luoni11}, and briefly described here in \app{torus}, provides an approximate representation of the global photospheric magnetic field distribution observed during the emergence of bipolar ARs. 
It consists of a half-torus field structure with uniform twist (both along and across its axis) that 
%in which the upper half of the torus 
is set to progressively emerge through the photosphere without distortion \citep[see {\fig{FR_sketch} and} the Appendix in ][]{Poisson16}. 
For its simplicity, the model does not include the deformations and reconnections occurring during its emergence, or any interaction between the magnetic field and the plasma. Still, it was shown that this model is a powerful and fast tool to represent the global properties of bipolar ARs \citep{Poisson15a, Poisson20a}. 

{The model is {strictly sub-photospheric} and its application is limited to reproduce only global aspects of the photospheric magnetic flux distribution of $\beta$-type ARs. More complex ARs ({\eg} those including $\delta$-spots) cannot be reproduced with a single toroidal FR defined by the half-torus model, unless the particular characteristics of the AR allow a separation in different individual bipoles. This work aims to analyze global aspect of simple bipolar ARs ($\beta$-type ARs), which correspond to the most commonly {observed ones} during any stage of the solar cycle \citep{Nikbakhsh19}.
} 

%{\S\bf --- About the model parameters}\\ 
The FR model construction is based on a set of free parameters. The { half-torus model}
has three dimensional and one non-dimensional parameters, as follows. The small radius $a$ and the large radius $R$ are associated with the size of the main polarities and the distance between their barycenters, respectively. The maximum field strength (at the axis of the torus) is defined by $B_0$ in Gaussian units. 
The sign and amount of magnetic twist is given by imposing the non-dimensional parameter $\Nt$, or twist number, corresponding to the number of turns of magnetic field lines around half of the torus axis \citep[see Figure 2 in][]{Luoni11}. {For a better understanding of these parameters, we add a sketch in \app{torus} together with the equations of the axial and azimuthal field components of the model.}

 %{\S\bf --- Synthetic magnetograms}\\
In order to produce synthetic magnetograms we cut the half torus with successive planes, each of which corresponds to the relative position of the photospheric plane as the FR emerges. We define the $z$ axis as the normal to the photospheric planes. For each particular magnetogram, $z=0$ corresponds to the location of the photospheric plane.
The torus axis of revolution (i.e., the half-torus base) is set at a depth $z=-d$ (see \fig{FR_sketch} in \app{torus}).
The computation of the field component $\Bz$ at $z=0$, for different values of $d$, produces a series of magnetograms that are determined by the values of the five model parameters: ($a$,$R$,$\Nt$,$B_0$,$d$).  
Moreover, since the model applies to observed ARs, we need three additional parameters to define the position and the orientation of the FR within the magnetogram. Therefore, we include the parameters $\xc$ and $\yc$, which indicate the mean point between the main polarities in Cartesian coordinates in the magnetogram frame ($x$ and $y$ for the east-west and the south-north directions, respectively). The last parameter is the FR tilt $\phi$, defined as the inclination of the polarities with respect to the east-west direction.
Finally, we define $\Mpi$ as the synthetic magnetogram obtained with the set of parameters $\vp_i = [a^{(i)},R^{(i)},\Nt^{(i)},\Bo^{(i)},d^{(i)},\xc^{(i)},\yc^{(i)},\phi^{(i)}]$, where the index $i$ corresponds to {a given set of parameters}.

%{\S\bf --- normalized depth}\\
The depth $d$ can be normalized using the torus size as follows. We use the apex distance, $R+a$, to introduce the {non-dimensional} parameter $d_0$ as:
  \BE \label{eq_dcorr}
   d = (1-d_0)\, (R+a) \,.
  \EE
\noindent Here the parameter $d_0$ ranges between 0 and 1, and corresponds to the fraction of the FR that has already emerged. 

%%%%%%%%%%%%%%%%%%%%%%%%%%%%%%%%%%%%%%%%%%%%%%%%%%%%%%%%%%
\section{Data modeling} \label{sect_modeling}

In this section we summarize the main characteristics of the modeling approach. The inference scheme was implemented using the open source library PyMC3, which provides all the needed tools for probabilistic analysis \citep[e.g., sampling and variational fitting algorithms,][]{PyMC3}.  PyMC3 relies on the Theano package for tensor algebra support, automatic differentiation, optimization and dynamic C compilation.

\subsection{Bayesian probabilistic analysis} \label{sect_bayes}

% {\S}{\bf --- Posterior probability} \\ 
The aim of the Bayes inference scheme is to obtain an estimation of the posterior probability  $P(\vp|\Mo)$, that provides the probability with which a certain combination of the model parameters $\vp$ replicates the observed magnetogram $\Mo$. 
We can obtain the posterior probability distribution, or simply, the ``posterior'', using the Bayes theorem expression:
  \BE \label{eq_Post}
   P(\vp|\Mo)  =  \frac{\cL(\Mo;\Mpi,\sigma) \, P(\vp)}{P(\Mo)}  \,,
  \EE
\noindent where $\cL(\Mo;\Mpi,\sigma)$ %$P(\Mo|\vp)$ 
is the conditional probability or likelihood (where the parameter $\sigma$ is the standard deviation between the model and the observations, defined later on \sect{likelihood}), $P(\vp)$ defines the prior probability distribution of model parameters (or simply, the ``prior''), and $P(\Mo)$ is the model evidence or marginal likelihood. %, respectively. 

% {\S}{\bf --- Marginal posteriors} \\ 
The posterior, $P(\vp|\Mo)$, for our model is an 8-dimensional distribution in which covariance elements are present. However, given the nature of the present problem, statistical independence between model parameters is assumed for this first application (the removal of this hypothesis is left for a future development).
Therefore, we are not considering a priori correlations between the parameters (first order approximation of $\cL$) and we limit our analysis to the marginal posteriors.  The marginal posterior for each parameter is obtained by integrating the full posterior in \eq{Post} over the remaining 7 parameters within the prior bounds. Therefore, each parameter will have a marginal posterior distribution. 

% {\S}{\bf --- Evidence} \\ 
The model evidence, $P(\Mo)$ in \eq{Post}, is the same for all combinations of $\vp$, so this factor does not affect the relative probabilities of different parameters.
The evidence $P(\Mo)$ describes the probability distribution of observing $\Mo$ with all the possible parameters $\vp_i$ as defined with the prior probabilities. This factor is useful for model comparisons but it is generally costly to derive as it requires to numerically integrate the whole probability distributions.

% {\S}{\bf --- Prior probability} \\ 
$P(\vp)$ introduces into the inference the information we have about the model parameters. In this sense, the prior comprises our knowledge of the physical system we aim to model (i.e., we guarantee that some values are correctly defined in specific physically reasonable ranges). 
Despite the relevance of the prior in \eq{Post}, we will see that if the model provides a good approximation to the observations, the information provided by the prior will be washed away from the posterior distribution.

% {\S}{\bf --- Defining the likelihood function} \\ 
In a Bayesian framework, the likelihood function $\cL$ models the probability of the observed data given the set of model parameters.  Assuming that the magnetogram is well calibrated, i.e. without bias, and characterized by an observational error $\sigma$, the maximum entropy distribution for the likelihood is a normal or Gaussian distribution $N(0, \sigma)$,
where the mean is zero since we assume the absence of systematic errors, {and $\sigma$ is the standard deviation}.

Then, for an observed magnetogram $\Mo$ with field values $\Bj{o}$ and a model magnetogram $\Mpi$ with field values $\Bj{i}$ computed with the set $i$ of parameters $\vp_i$, the likelihood function is defined as
  \BE \label{eq_likelihood}
  \cL(\Mo;\Mpi,\sigma) = \frac{1}{\sigma\sqrt{2\pi}} 
            \exp \Bigg(-\frac{1}{2 \sigma^2} \sum_j (\Bj{o}-\Bj{i})^2 \Bigg) \,,
  \EE
where $j$ {is indexing} the magnetogram pixels, then the summation over $j$ indicates a summation over all these pixels.  The exponential expression in \eq{likelihood} can take extremely low and large values, therefore the PyMC3 tools use the logarithm of $\cL$ to avoid truncation errors.

The posterior distribution obtained will be conditioned by the selection of $\cL$, in this case a Gaussian-like likelihood. { This will produce the} same function-like distribution for the posterior. The selection of a normal distribution is accompanied by the hypothesis that the error between the model and the observation is due to an aleatory process and, as a consequence of this selection, the width of the posterior is directly proportional to $\sigma$. Still, a second order deviance can be present on the posterior due to the correlation between parameters which affects the symmetry of the distribution.

%  {\S}{\bf --- Selecting a prior distribution} \\ 
For simplicity we assume that no a priori information can be obtained from the magnetograms, so we explore a broad parameter space defined by uniform prior distributions (except for $\Nt$ {as explained below}). In this sense each prior distribution is defined by two values, the lower and the higher limits. This kind of distribution assigns equal probability within the defined ranges and zero outside this interval.
We use the same selection of prior ranges for all the magnetograms.  

%  {\S}{\bf --- Prior for $\Nt$} \\ 
In \citet{Poisson15a}, we obtained the distribution of $\Nt$ values computed from the polarity inversion line (PIL) inclination of 41 ARs. The distribution has a strong skewness towards small values, with a mean of $0.3$ and a standard deviation of $0.2$. These results were extended over 168 ARs in \citet{Poisson16} with similar statistical conclusions.  
Therefore, for the prior of $|\Nt|$ we use a Gamma distribution with {the above} mean and standard deviation values. This distribution is defined strictly positive, meaning that we allow only one twist sign for the FR and we do not limit the upper-bound for the twist. This {Gamma distribution} only reduces the probability of having larger values of $|\Nt|$, which are not observed in ARs. Additionally, the information of the twist sign has to be set using the criteria presented in \citet{Luoni11}, which rely on the tongue pattern that is independent of the polarity signs and require the computation of the acute angle formed by the PIL and the bipole axis. In the case of AR 10268, we found a negative twist sign \citep{Poisson15b}.

%%%%FIGURE posterior %%%%%%
\begin{figure}[!t]
\centering
\includegraphics[width=0.45\textwidth]{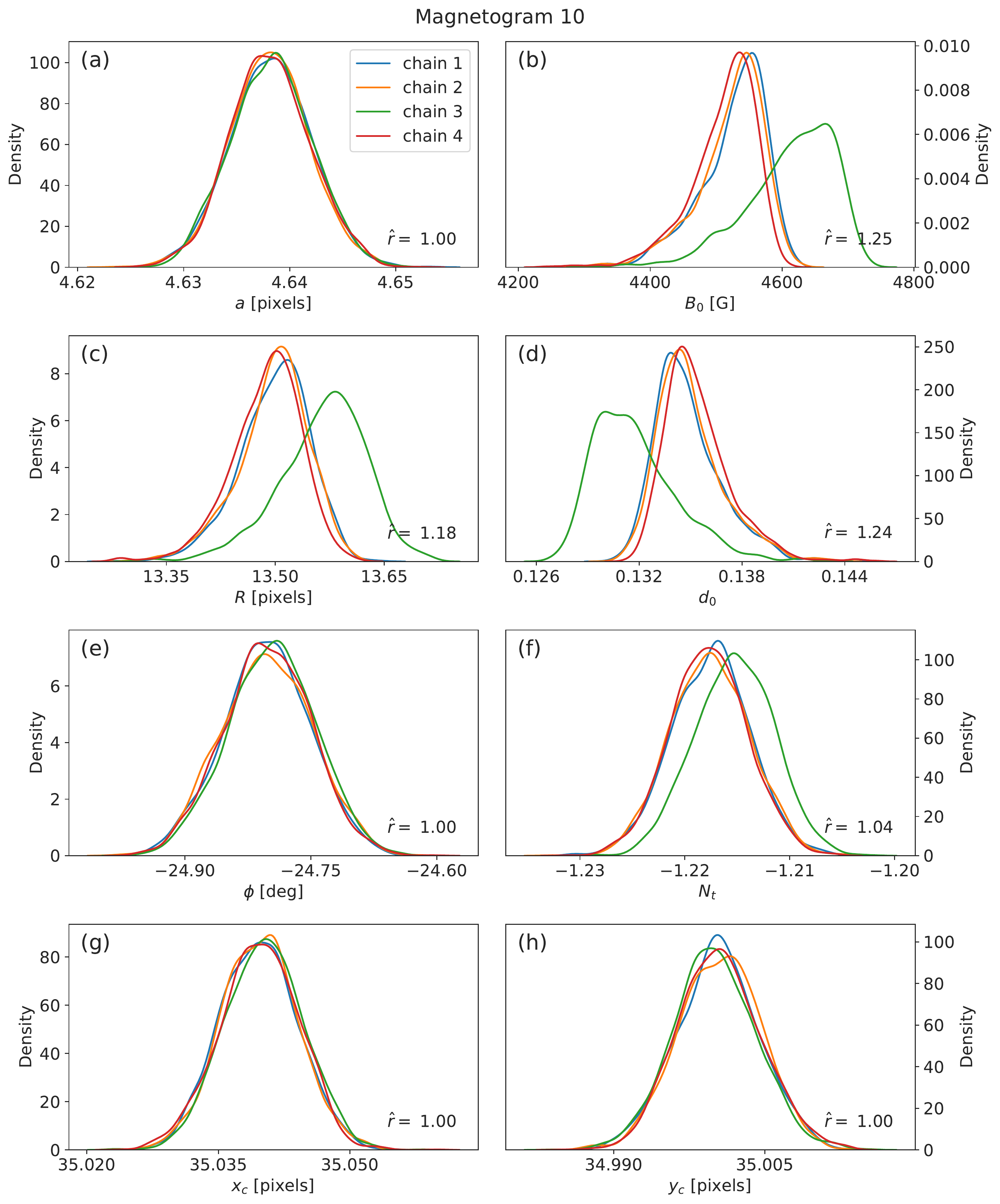}
\caption{Marginal posteriors obtained for the model parameters of magnetogram number 10 of AR 10268 (see \fig{AR10268}b). Colors correspond to the four different chains (indicated in the inset of panel a) having 2000 samples each. The panels correspond to the parameters: (a) $a$ , (b) $B_0$, (c) $R$, (d) $d_0$, (e) $\phi$, (f) $\Nt$,  (g) $x_c$, and (h) $y_c$.
{The $\hat{r}$ value is written at the bottom left of each  panel.} $\hat{r}$ is the normalized rank factor used to quantify the convergence between different MC Markov chains to a single distribution. $\hat{r}$ tends to unity when convergence is achieved.
}
\label{fig_posterior}
\end{figure}
%%%%FIGURE posterior %%%%%%

\subsection{Sampling the parameter space}
\label{sect_sampling}

%  {\S}{\bf --- Not just fitting but modeling} \\   
Finding the parameters of maximum likelihood entails an optimization problem that can be solved using several methods, e.g. Powell's conjugate direction method among them. Most of these methods are fast but can be trapped in local maxima depending on the selected initial point. In particular, they provide the location of the maxima of $\cL$, being the path by itself of little interest and how $\cL$ behaves around its maximum value not considered.
Therefore, most of these methods give no information on the precision of the results, unless the second order derivatives of \mp{$\cL$} are also evaluated. {In contrast,} the formalism given by \eq{Post} presents an statistical approach in which a probability distribution can be sampled, allowing a deeper analysis of the model parameters (e.g., mean, median and variance).  But this approach needs specific algorithms to efficiently explore the parameter space  to reconstruct $P(\vp|\Mo)$ in a feasible number of iterations. 

%  {\S}{\bf --- Exploring the surrounding of the maximum} \\
Monte Carlo methods provide {a more comprehensive description of the posterior distribution}, not just single point estimates, by producing a set of random samples within the defined parameter space.  More generally, Markov Chain Monte Carlo \citep[MCMC,][]{Neal93} algorithms provide a more efficient way to explore broad parameter spaces with less computations. Nevertheless, sampling distributions with multiple peaks cannot be completely described using general {common} MC algorithms, as they often get stuck in one of the secondary maxima (e.g., as with Hamiltonian Monte Carlo methods). 

Since our model is complex and non-linear, the posterior can in principle have multiple secondary maxima.  Hence, we implement a Sequential Monte Carlo (SMC) sampler \citep[see e.g., ][]{KANTAS2009}. This sampler, included in the PyMC3 package, is a well established Monte Carlo algorithm based on a Simulated Annealing approach combining important sampling, tempering, and MCMC kernels \citep{CHOPIN2020}. In this case, the posterior is expressed as: 
  \BE \label{eq_Post_beta}
   P(\vp|\Mo)_\beta  \propto  {{(\cL(\Mo;\Mpi,\sigma))}^\beta \, P(\vp)}  \,,
  \EE
\noindent where $\beta$ is an auxiliary temperature{-like} parameter to control the sampling sequences. In the zero stage, $\beta=0$, the algorithm initializes directly sampling the priors. PyMC3 optimizes this sampler by setting a number of steps in which the value of $\beta$ is progressively increased. Every two successive $\beta$ steps an importance sampling stage is initialized. In each stage, weights are computed as the ratio of the tempered posterior of the respective sample between two successive $\beta$ steps. These weighted samples are then used by an importance sampling algorithm with which new MCMC draws are obtained. Samples are then used as seed for the next $\beta$ step. Iteration is complete once $\beta$ reaches unity producing the final sample collection. 

We perform four independent chain samples, with 2000 samples each. Each chain has different random seeds for the initial points, meaning that no correlation is imposed between chains. 
\fig{posterior} shows an example of the marginal posterior for the model parameters obtained for magnetogram number 10 of AR 10268. The distributions are constructed with the Kernel Density Estimator \citep[KDE,][]{Parzen62}. In this case the {standard deviation} between model and observations ($\sigma$) is set to  be $10$ G, assuming the instrumental error is the only contribution to $\sigma$. Each color indicates different MC Markov chains.

%  {\S}{\bf ---  Convergence of posteriors} \\ 
The normal-like posterior distributions and the progressive convergence of consecutive chains indicate that optimal model parameters for the analyzed observed magnetogram can be obtained from the mean value of the posterior. 
Convergence is 
guaranteed for an infinite number of samples. In practice we cannot prove that the posterior obtained from a finite number of samples has converge to the real one. That is why it has been accepted that independent chains, with scattered initial points, converging to the same distribution is the best we can use to say we have achieved convergence.
The normalized rank, $\hat{r}$, diagnostic tests \citep{Vehtari2021} provides a way to quantify the degree of convergence by comparing the variance between multiple chains with the variance within each chain. If convergence has been achieved, the between-chain and within-chain variances should be identical, hence $\hat{r}=1$. Values of $\hat{r}$ above $1.4$ indicate problematic distributions and convergence is not guaranteed. In the present example, $\hat{r}$ values are below $1.25$, and often close to 1 (see $\hat{r}$ values in the panels of \fig{posterior}).

%%%%FIGURE magnetograms %%%%%%
\begin{figure}[!t]
\centering
\includegraphics[width=0.45\textwidth]{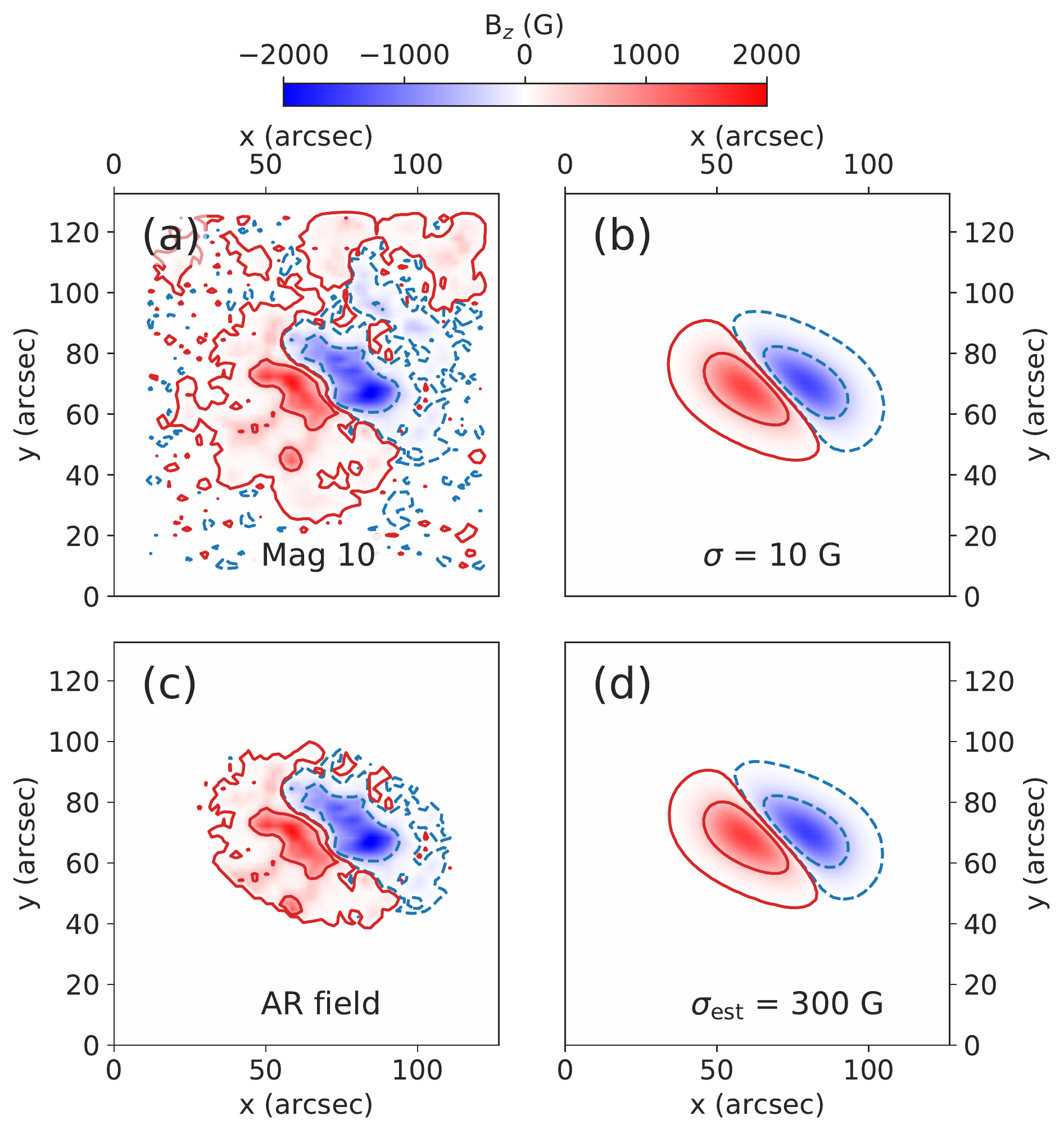}
\caption{ (a) Magnetogram {number} 10 for AR 10268. 
  (b) {Most probable} model for {the same magnetogram} obtained with $\sigma = 10$ G.
  (c) Magnetogram {number} 10 limited to the AR {field} defined {using} 
the {most probable} model {shown in} panel (b), fulfilling the condition $|B_{\rm +,j}| \geq 1 $ G. 
  (d) {Most probable} model for panel (c) data obtained with $\sigma = 300$ G. The {color and contour conventions are} the same as in \fig{AR10268}. {The movie mov-fig3.mp4, from the supplementary material, shows the magnetograms and their associated models during the full emergence.}  
}
\label{fig_mags}
\end{figure}
%%%%FIGURE magnetograms %%%%%%

%%%%FIGURE sigmas %%%%%%
\begin{figure}[!t]
\centering
\includegraphics[width=0.45\textwidth]{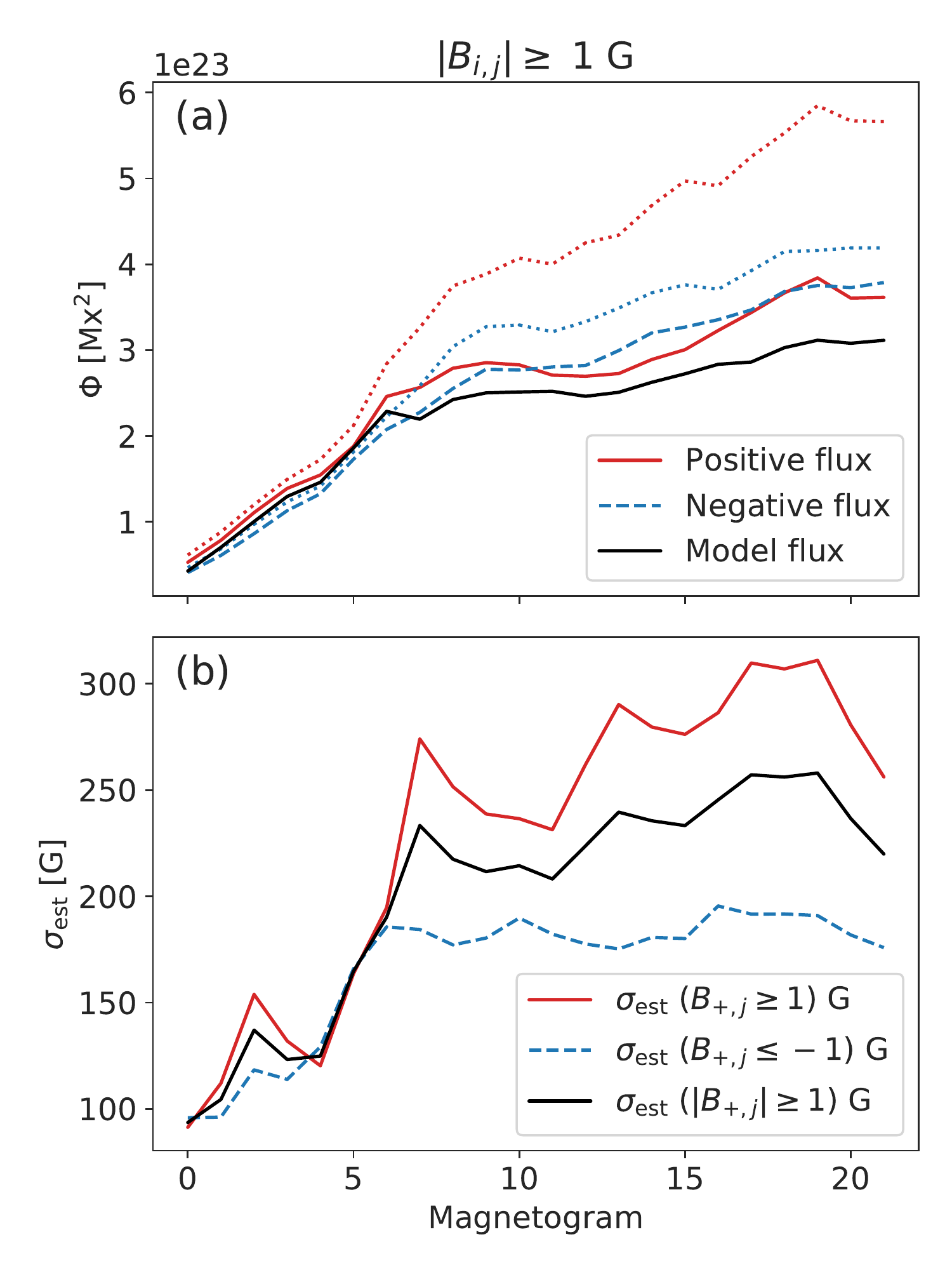}
\caption{
(a) Evolution of AR 10268 magnetic flux.  The dotted-red (blue) line shows the evolution of the positive (negative) magnetic flux of the square selection of the magnetogram  (as shown in \fig{AR10268}). Solid-red and dashed-blue lines are the positive and negative flux, respectively, computed over the AR field (see \fig{mags}c as example).  The black line shows the evolution of the most probable model magnetic flux (positive plus absolute value of the negative fluxes divided by two). 
(b) Estimated value for the {standard deviation} between model and observations using \eq{sigmaEstim} over the AR field. The
red-solid (blue-dashed) line corresponds to $\sigma_{\rm est}$ for the positive (negative) polarity. The black line corresponds to $\sigma_{\rm est}$ for points within the AR region ($|B_{\rm +,j}| \geq 1$ G).}
\label{fig_sigmas}
\end{figure}
%%%%FIGURE sigmas %%%%%%

\subsection{Inference error}
\label{sect_likelihood}

%{\S}{\bf --- The parameter $\sigma$ } \\ 
$\cL$ is a function of the parameter $\sigma$, which corresponds, in principle, to the standard instrumental deviation of the observed data. However, in practice the {standard deviation} between the model and the observations is unknown and larger than the instrumental error, because the model is not exact, as it represents, at best, only the global aspect of the flux distribution and does not reproduce the small scale variations correctly. In the case of MDI magnetograms the typical instrumental noise level is 9 G, while the standard deviation between model and observations is expected to be much larger (in view of the simplicity of the model). Therefore, even for the most probable fit, $\sigma$ is expected to be dominated by the difference between observations and model. For this reason, we propose to estimate its value based on the computation of the residuals between the most probable fitted model and the observed magnetograms. 

At first, we can obtain a {most probable} fit by conducting an inference assuming the ideal modeling, in this sense $\sigma$ is set to $10$ G.
But a more appropriate value can be found, once the {most probable} model is approximately known, by computing   
  \BE \label{eq_sigmaEstim}
  \sigma_{\rm est} =  \sqrt{ \frac{1}{N_d} \sum_j (\Bj{o}-B_{\rm +,j})^2 } \,,
  \EE
\noindent where $N_d$ is the number of pixels in $\sum_j$ and the subscript "+" indicates the {most probable} model with parameters $\vp_{\rm +}$ 
derived from the mean values of the marginal posteriors (e.g., \fig{posterior}).

%%%%FIGURE boxplot %%%%%%
\begin{figure*}[!t]
\centering
\includegraphics[width=0.9\textwidth]{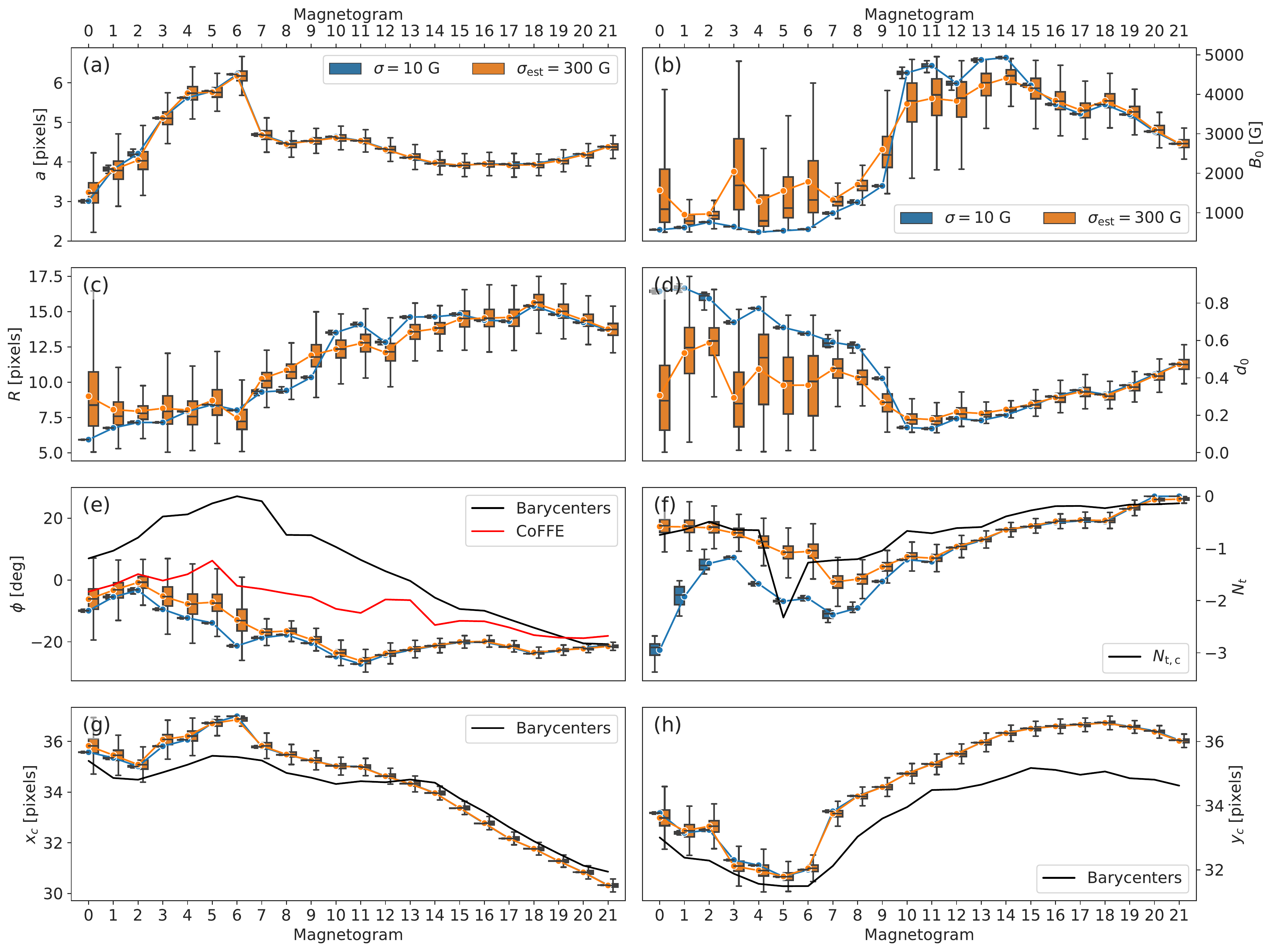}
\caption{Box plots showing the evolution of the marginal posteriors inferred for 22 magnetograms along the emergence of AR 10268.
Each panel shows the evolution of the model parameters: {(a) $a$ , (b) $B_0$, (c) $R$, (d) $d_0$, (e) $\phi$, (f) $\Nt$,  (g) $x_c$, and (h) $y_c$.} Blue and orange boxes compare the {marginal posteriors obtained using $\sigma = 10$ G and $\sigma_{\rm est} = 300$ G,} respectively. {The boxes represent the three-quartile values of the distribution {(so boxes contain the 50\% central part of the distribution)} and the {whisker extensions are} fixed by a factor $1.5$ of the inter-quartile range.} 
The blue and orange lines mark the evolution of {the mean value of the marginal posteriors for either $\sigma$ or $\sigma_{\rm est}$}. In panel (e) the tilt {values obtained using two different methods} are added (see inset). {The black line in panel (f) corresponds to the estimation of $N_{\rm t,c}$ obtained comparing the inclination of the PIL with respect to the direction of the AR bipole}. Black lines in panels (g) and (h) correspond to the AR barycenter. 
}
\label{fig_boxplot}
\end{figure*}
%%%%FIGURE boxplot %%%%%%

%%%%%%%%%%%%%%%%%%%%%%%%%%%%%%%%%%%%%%%%%%%%%%%%%%%%%%%%%%
\subsection{Model of AR 10268}    \label{sect_ModelAR}

%  {\S}{\bf --- Best model with $\sigma = 10$ G} \\
{As mentioned in the previous section, a} first estimation of the most probable parameters of the model is obtained using $\sigma=10$ G. The posterior obtained is extremely narrow compared with the bandwidth defined for the priors (see \fig{posterior}), so the real uncertainties of the parameters (posterior variance) are not well represented. Still, the mean of each marginal posterior can be used to compute the most probable fit to the observed magnetograms.

\fig{mags}b shows the most probable model of magnetogram number 10 using $\sigma = 10$ G. Comparing it with the observations, we see a good agreement with the shape and size of the magnetic tongues, the inclination of the PIL, the tilt angle and the field strength in the core of each polarity (inner contours). Still, differences are evident by looking at the small scales and lower field pixels surrounding the AR core. 

 % {\S}{\bf --- Filtering background field } \\ 
To remove the contribution of the background field, i.e. not belonging to the AR, we reduce the magnetogram to those pixels in which $|B_{\rm +,j}| \geq 1$ G. 
\fig{mags}c shows this restricted magnetogram, henceforth called simply AR field. 
This allows us to derive parameters for the model which are more representative of the studied real AR field %{\bf No ser\'\i a mejor poner real AR field?}
(i.e., the FR intrinsic field).
\fig{sigmas}a shows the evolution of the AR positive magnetic flux, with a solid-red line and the negative flux with a dashed-blue line. The AR field selection significantly reduces the flux unbalance due to the contribution of the background positive field.

%  {\S}{\bf --- Estimating $\sigma$} \\ 
The value of $\sigma$ needs to incorporate the fact that the model cannot reproduce all the complex features of the AR emergence, and therefore we perform a second inference in which a more realistic value is used. As a consequence of this new estimated value, we will obtain %different (
{broader} posterior distributions from which the uncertainty of the parameters can be estimated. 
Comparing the most probable model magnetograms with the AR field we obtain an estimation of {what we now call $\sigma_{\rm est}$} for each polarity using \eq{sigmaEstim}. \fig{sigmas}b shows how this estimation varies during the emergence. We find an expected asymmetry \citep[\eg ][]{Lites98} between the leading (negative) and following (positive) polarities. This indicates not only the limitation of the model to fit the more disperse flux of the following polarity, but also that the existence of this asymmetry introduces greater uncertainties on the parameters since the half-torus model is strictly symmetric.  In this sense the model will try to fit both polarities with the same parameters despite their differences.  A development of the model to take into account this asymmetry is postponed for a future study.

%  {\S}{\bf --- Second estimation mag} \\ 
In the present set up and the testing of the statistical method, we select 
the maximum value of $\sigma_{\rm est}$ in \fig{sigmas}b. This introduces an upper bound for the statistical errors, especially in the earlier phase of emergence. This uniform value allows us to test the sensitivity of the  parameters of the model during the emergence phase from the broadness of the posteriors. Therefore, our second inference run is done over the AR {field} using {$\sigma_{\rm est} = 300$ G (the largest value} achieved during the evolution). \fig{mags}d shows the best magnetogram obtained with this second inference. 
There are no clear differences between panels (b) and (d), implying that the first approach is good enough to mimic the field distribution of the AR. Still, there are significant differences between the posterior distributions obtained with both estimations. They are analyzed below.

%  {\S}{\bf --- evolution of parameters} \\ 
We extend the above analysis of the posteriors to 22 magnetograms covering the full emergence of AR 10268. The movie mov-fig3.mp4 in the supplementary material includes the evolution of the magnetograms for the AR 10268, the AR field, and the respective models. We model each magnetogram individually without imposing any temporal correlation on the parameters. In \fig{boxplot} we plot the evolution of the marginal posteriors using box and whisker plots. The segments in the boxes indicate the mean values. We compare models obtained with the first inference procedure ($\sigma= 10$ G, blue), and second inference ($\sigma_{\rm est} =300$ G, orange) including background filtering. Some of the peaks of the first inference posterior are so narrow that the corresponding boxes are not {visible} in the plots.

%  {\S}{\bf --- first half} \\ 
The first part of the emergence (up to magnetogram number 10) is characterized by a broader distribution in some parameters, especially $d_0$, which presents the greatest dispersion. The limited information provided by the magnetogram pixels at the beginning of the emergence is expected to produce broader posteriors induced by a stronger correlation between the parameters of the model. In particular, different combinations of the parameters $d_0$, $R$, and $\Bo$ can produce similar synthetic magnetograms, e.g. a FR almost fully emerged with low axial field and a FR deep down the photosphere with large axial field can produce similar field distributions with $\sigma_{\rm est} = 300$ G. 

%  {\S}{\bf --- second half} \\
In the second half, {the marginal posteriors for most of the parameters} become narrow and the mean value of both estimations ($\sigma = 10$ G and $\sigma_{\rm est} = 300$ G) converge.  At that point the model is well constrained by observations.  Nevertheless, $\Bo$ is the most sensitive parameter, having the greatest dispersion up to the end of the emergence. 

%  {\S}{\bf --- agreement of some model parameters} \\ 
For most of the analyzed magnetograms, the parameters $a$, $\phi$, $\xc$, and $\yc$ converge to similar values independently of taking {$\sigma$ or $\sigma_{\rm est}$}.  This implies that along the direction of these parameters the $\cL$ function has a narrow and steep main peak.  Moreover, the plot of the posterior (as in \fig{posterior}) indicates that 
other secondary peaks are shallow.

%  {\S}{\bf --- number of turns} \\ 
\fig{boxplot}f shows the estimation of $\Nt$. {The values obtained with $\sigma = 10$ G and $\sigma_{\rm est}= 300$ G depart during} the first half of the emergence indicating a possible anti-correlation with $\dnot$. In this sense the elongation of a polarity can be represented by the number of turns but also by the stage of the emergence, i.e. increasing $\dnot$ above $0.6$ reduces the {extension of the tongues} despite the large values of $|\Nt|$ {(see the difference between the orange and blue boxes in panels (d) and (f) until magnetogram number 8)}. The evolution of $\Nt$ is associated to the variation of the polarity elongations. 
The maximum elongation of the positive polarity is reached around the maximum value of $\Nt$ (see movie mov-fig3.mp4, or Figure 6d in \citet{Poisson15a}). Moreover, $\Nt$ can be correctly estimated only when the elongation of the polarities is significant, therefore its determination is limited along the emergence, i.e. $\Nt$ tends to zero once the tongues have retracted. Finally, the estimation obtained with $\sigma_{\rm est} = 300$ G is in good agreement with the parameter $N_{\rm t,c}$ (black line in \fig{boxplot}f) obtained for this AR in \citet{Poisson15a} with a different method (i.e., using the relative orientation of the PIL with respect to the direction of the AR bipole axis). 

%  {\S}{\bf --- tilt} \\ 
The tilt angle obtained from the inference (\fig{boxplot}e) indicates a sustained $20\degree$ clockwise rotation. This estimation of $\phi$ provides a correction of the effect of the magnetic tongues on its determination using the magnetic barycenters (black line).  The difference between both  methods proves the significant effect of the tongues over this quantity during the emergence.
We see that the correction provided by CoFFE \citep{Poisson20a} partially improves the estimation of $\phi$.

%  {\S}{\bf --- AR center} \\ 
The evolution of {the AR center,} $x_c$ and $y_c$, indicates a drift of AR towards the east-north direction.  {This drift is} towards the north pole since AR 10268 is located in the northern hemisphere ($12^{o}$) (see \fig{boxplot}g,h).  This northern drift is plausibly associated with the Sun's meridional flow. 
The AR center can also be estimated from the computation of the barycenters of the main polarities (black line).  This estimation is in good agreement with the obtained $x_c$ and $y_c$, with differences within the magnetogram spatial resolution.

%%%%FIGURE modgen %%%%%%
\begin{figure}[!t]
\centering
\includegraphics[width=0.45\textwidth]{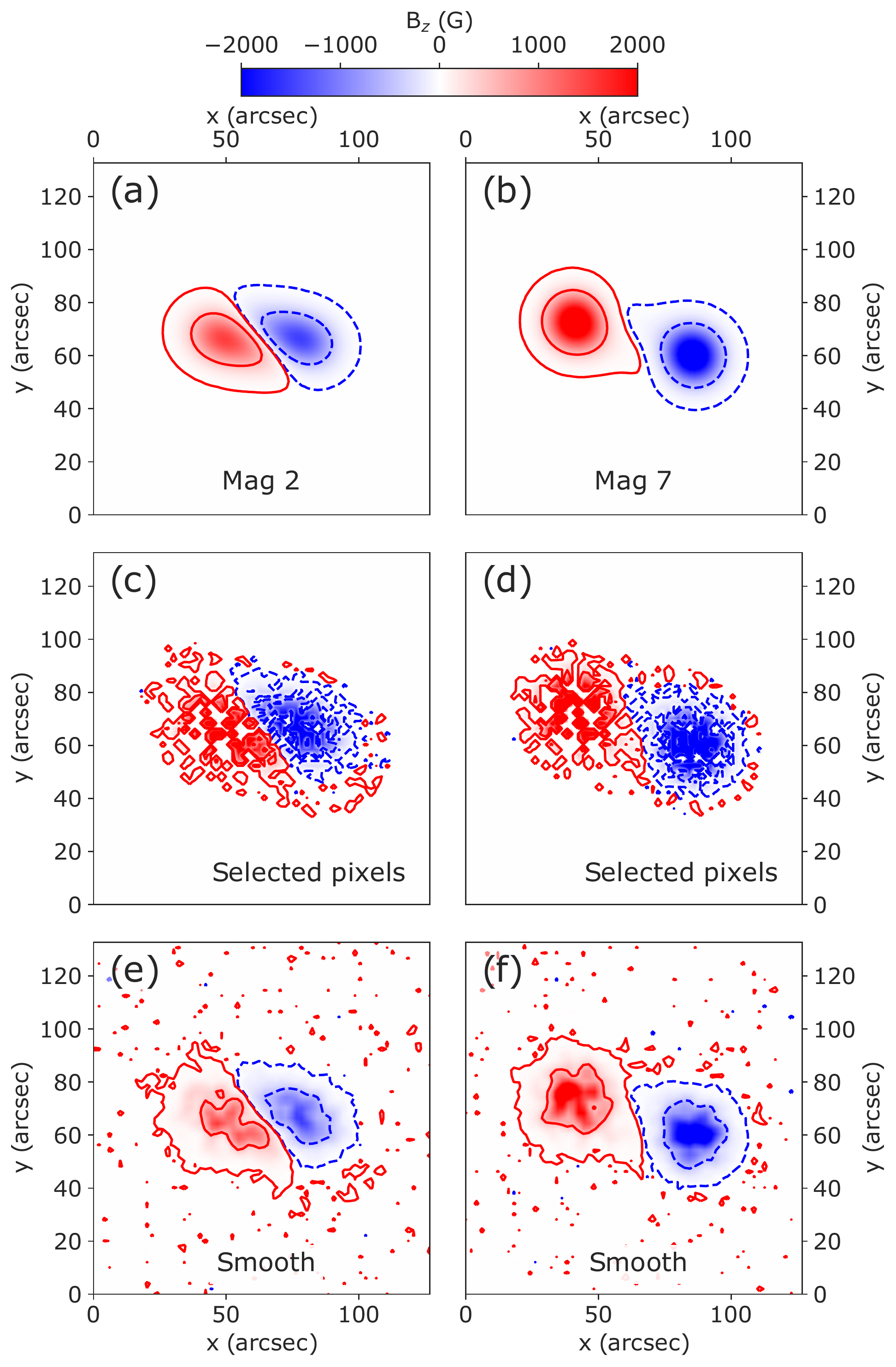}
\caption{Example of the model generator: (a)-(b) Magnetogram 2 and 7 of the seed model ($M_0$). (c)-(d) Random selection of pixels for each magnetogram ($M_2$).
(e)-(f) Simulated data generated from panels (c) and (d), after applying a Gaussian-smooth filter and normal noise ($M_4$).
The color convention is the same as in \fig{AR10268}.
}
\label{fig_modgen}
\end{figure}
%%%%FIGURE modgen %%%%%%

%%%%FIGURE models boxplot %%%%%%
\begin{figure*}[!t]
\centering
\includegraphics[width=0.9\textwidth]{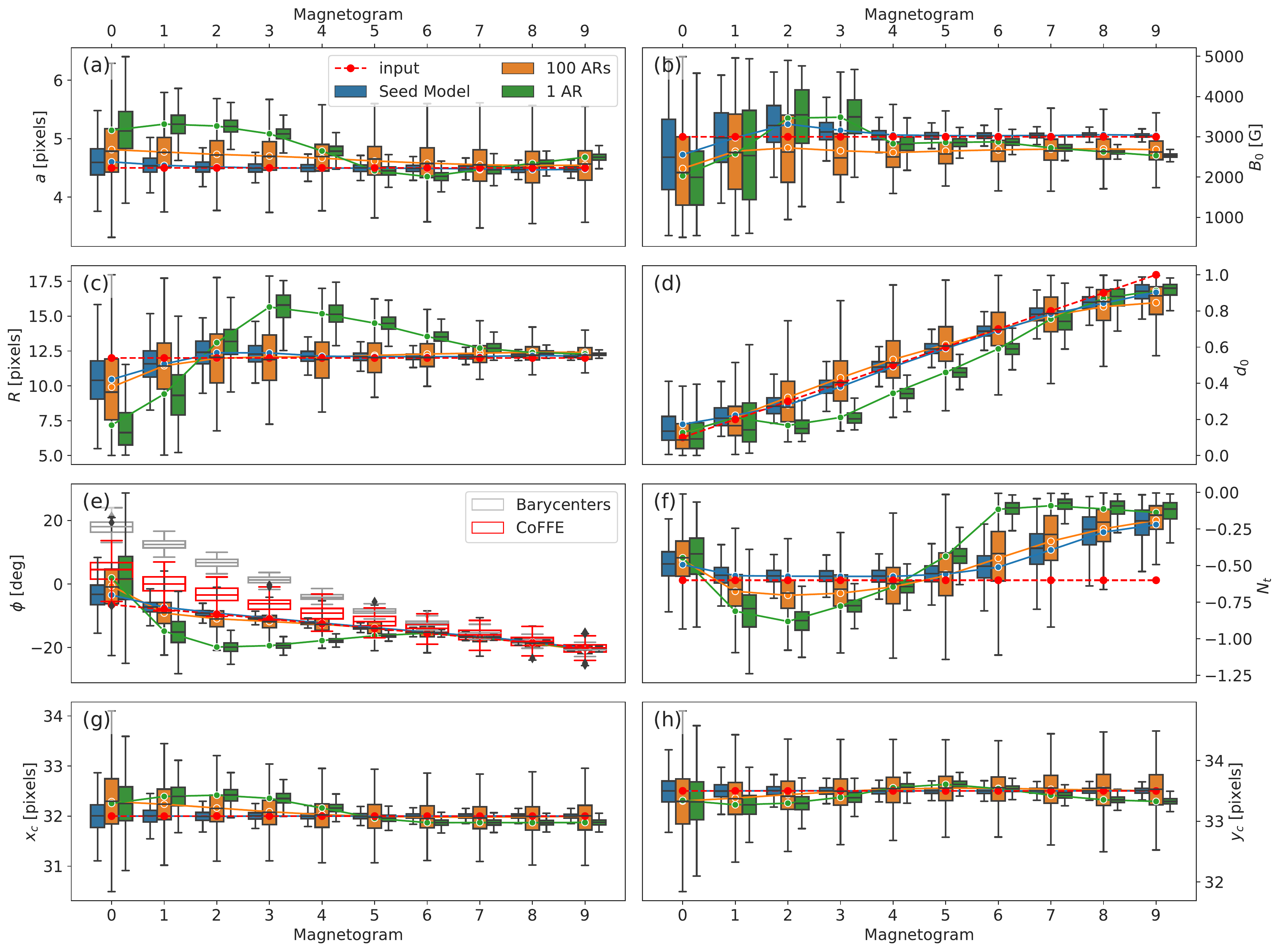}
\caption{Box plots showing the evolution of sampled marginal posteriors for the parameters of the seed model (blue), 100 generated ARs (orange), and one individual AR as example (green). For the orange boxes, each marginal posterior is constructed by 200000 samples concatenated from the 2000 samples for each of the 100 generated ARs. 
Each panel shows the evolution of the model parameters: (a) $a$ , (b) $B_0$, (c) $R$, (d) $d_0$, (e) $\phi$, (f) $\Nt$,  (g) $x_c$, and (h) $y_c$.
The inference is made assuming $\sigma = 300$ G. The blue, orange, and green lines show the evolution of the corresponding marginal posterior means. The red-dashed line shows the seed model input parameters.
The gray and red boxes of panel (e) show the evolution of the tilt angle obtained using the magnetic barycenters and CoFFE for 100 generated ARs, respectively.
}
\label{fig_modbox}
\end{figure*}
%%%%FIGURE models boxplot %%%%%%

%%%%%%%%%%%%%%%%%%%%%%%%%%%%%%%%%%%%%%%%%%%%%%%%%%%%%%%%%%
\section{Effect of perturbations}   \label{sect_modelgen}

In this section we test the statistical method by applying it to a series of magnetograms obtained with the FR emergence model (i.e., synthetic magnetograms). We also study the effect of including a series of non-linear perturbations to the synthetic magnetograms to mimic small scale magnetic structures, as well as asymmetries observed between the polarities of the ARs, without the need of including complex interactions between the magnetic field and the plasma.

\subsection{Simple test} \label{sect_simplecase}

% {\S}{\bf --- Test model} \\
Using the half-torus model we generate a cube of 10 magnetograms (indexed from 0 to 9) to which we apply the already developed statistical method. This number of magnetograms provides enough resolution for the simulated AR emergence. 
{The obtained magnetograms cover the same field of view and spatial resolution as the observations of AR 10268, so simulated data has the same pixel size as the MDI instrument.
We tested the inference dependence on the spatial resolution (not shown here) reducing by a factor two and a factor three the original resolution of the MDI magnetograms for AR 10268. For both tests, we find no significant differences with the results shown in \fig{boxplot}.}

The emergence is modeled by the increase of $\dnot$ from $0.1$ to $1$. The intrinsic parameters of the torus, $a$, $R$, $\Bo$, $\Nt$, $\xc$ and $\yc$, are set as constants during the emergence, for simplicity ($a =4.5$ pixels, $R=12$ pixels, $\Bo = 3000$ G, $\Nt=-0.6$, $\xc=32$ pixels and $\yc=33.5$ pixels). The tilt angle $\phi$ linearly evolves with a clockwise rotation of $15^{o}$ (within a range of $-5^{o}$ to $-20^{o}$) which is typically observed in some emerging bipolar ARs \citep{Poisson15a}, including AR 10268. 
Two steps of the simulated emergence, corresponding to magnetograms (Mag) 2 and 7, are shown in \fig{modgen}a,b. The magnetic tongues are clearly visible in Mag 2 and more retracted in Mag 7.  The results of the inference procedure are shown in \fig{modbox} in the same format as in \fig{boxplot}. The marginal posteriors obtained for the test model are indicated with blue boxes and whiskers, while the red-dashed lines correspond to the input parameters of the model. Results shown with other colors are described in \sect{Inference}. 

%{\S}{\bf --- inference over the test model} \\
The marginal posteriors for the parameters of each magnetogram are Gaussian-like distributions (not shown, {but compatible with the symmetry observed in the {quartile boxes}). In \fig{modbox}, the marginal posteriors have a} mean value (blue lines)  very close to the half-torus model input parameters (red lines in \fig{modbox}). This basic test shows that the sampler algorithm works fine to recover the original parameters. 
The broadness of these posteriors change during emergence.  Since we set the value of the {standard deviation} between the model and the observations (called $\sigma$ from now on) to a constant that we take to be $300$ G, as our previously computed $\sigma_{\rm est}$, the change in the broadness {includes} the effect of overestimating $\sigma$ on the sampled posterior during the early phases of emergence (in particular, for parameters $\Bo$ and $R$). 
In future applications of the method to observed ARs, the expected lower values of $\sigma$ in this early phase (as in \fig{sigmas}b) will attenuate this effect.

%  {\S}{\bf --- Case of $\Nt$} \\
The broadness of the marginal posteriors decreases along the simulated emergence for all the parameters except $\Nt$.
Indeed, this is the only parameter which fully depends on the extension of the magnetic tongues (for other parameters the presence of the tongues rather perturbs their determination). Since tongues tend to disappear by the time the half-torus has emerged, the information of $\Nt$ reflected in the magnetogram also decreases during the emergence.  Furthermore, not only the marginal posterior of $\Nt$ for each magnetogram will become broader, but also its mean value will tend to depart from the input parameters. This indicates that $\Nt$ is not well constrained after $70\%$ ($\dnot \geq$ 0.7) of the total FR has emerged.

\subsection{AR generator} \label{sect_argen}

%  {\S}{\bf --- seed model} \\
In this section we develop an AR generator to simulate a large variety of small scale magnetic features, comparable to the ones observed in AR 10268 (see \fig{AR10268}), in order to explore how perturbations affect the posterior. For that we follow a four-step procedure that perturbs the magnetograms of our previous test model (\fig{modgen}a-b), which we then use as a seed for the AR generator.

%  {\S}{\bf --- 1est step} \\
In the first step the AR generator makes a random selection ($\mathbb{S}$) over a fraction of the pixels that comply $|B(x_j,y_j)|\ge 1$ G.
We choose a different number for the positive ($f_{\rm p}$) and negative ($f_{\rm n}$) flux pixels to mimic the observed asymmetry between the leading and the following polarities.
For smaller values of $f_{\rm p}$, the positive polarity becomes less homogeneous.  Selected pixel coordinates $(x_j,y_j)$ are obtained for each magnetogram of the cube.
The output of the first step, $M_1$, is defined by the condition $B_{\bar{\mathbb{S}}}= 0$ outside the pixel selection.

%  {\S}{\bf --- 2nd step} \\
In the second step, we multiply each pixel of the ${\mathbb{S}}$ selection by a constant factor in order to maintain the total magnetic flux of the original seed model. %P , i.e. $\Phi(M_0) = \Phi(M_1)$.  
We compute a factor for each polarity to make sure that the magnetic flux balance is conserved (in this step). In other words, we redistribute the flux in magnetic nodes where the flux is more concentrated producing a new version of the magnetogram cube, $M_2$ (see \fig{modgen}c-d). 

%  {\S}{\bf --- 3rd step} \\
In a third step, {we apply a Gaussian smoothing filter of 1 pixel width to the magnetograms}. The aim is to smooth discontinuities in the field distribution and produce more realistic magnetograms (called $M_3$). 
  
%  {\S}{\bf --- 4th step} \\
Finally, the fourth step introduces noise and flux unbalance to $M_3$, adding a random normal noise with a mean of $5$ G  and standard deviation of $10$ G to each pixel. Two examples of the final simulated magnetograms $M_4$ are shown in \fig{modgen}e,f.

\begin{figure}[!t]
\centering
\includegraphics[width=0.45\textwidth]{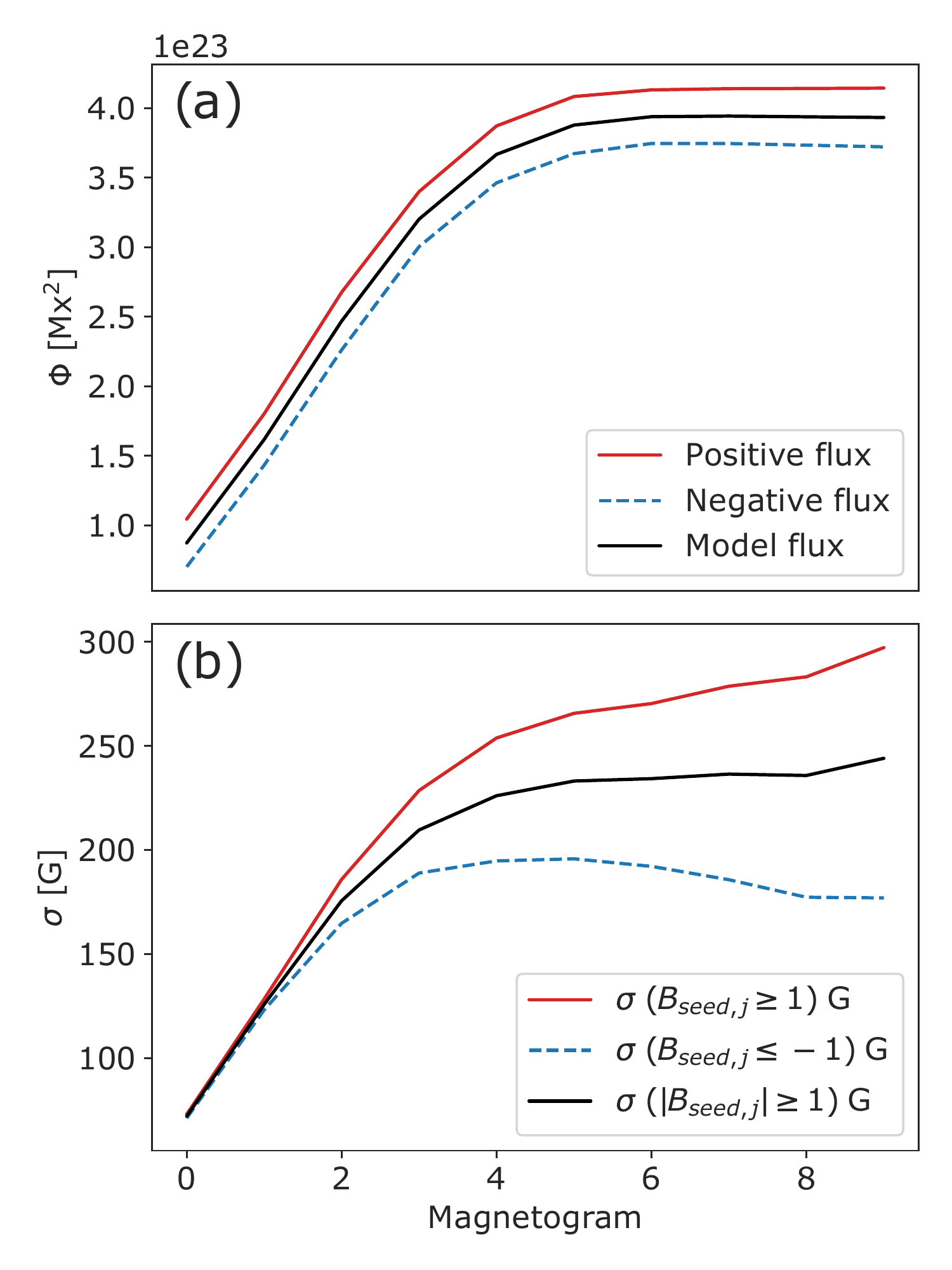}
\caption{(a) Evolution of magnetic flux for the generated AR shown in \fig{modgen}e,f. Solid-red and dashed-blue lines are the positive and negative flux, respectively. The black line shows the seed model magnetic flux. {(b) {Estimated $\sigma$ using \eq{sigmaEstim} {applied to} 
the seed model and the generated AR.}}
The red-solid (blue-dashed) line corresponds to $\sigma$ for the positive (negative) polarity. {The black line corresponds to $\sigma$ estimated with all the pixels within the generated AR {field}.} 
}
\label{fig_modsig}
\end{figure}

%  {\S}{\bf --- Description of \fig{modsig}} \\
The evolution of the positive (red-solid line) and negative (blue-dashed line) magnetic flux of a generated AR (M4) is shown in \fig{modsig}a. 
The excess of positive flux emulates the contribution of the background field as in AR 10268.  The black line corresponds to the flux of the seed model, $M_0$.  Its difference with the blue and red lines characterizes the imposed flux unbalance.
  
%  {\S}{\bf --- Generator parameters} \\
$f_{\rm p}$ and $f_{\rm n}$ are the main parameters of the AR generator, since they produce strong field perturbations that determine how far the output will be from the seed model (i.e. the value of $\sigma$).  If we consider that the seed model is the most probable model for the generated AR, then we can compute $\sigma_{est}$ from \eq{sigmaEstim} comparing the magnetograms of the generated AR ($M_4$) and the seed model ($M_0$). Therefore,  a selection of $f_{\rm p}$ and $f_{\rm n}$ will determine how $\sigma_{est}$ evolves during the emergence. We find that using $f_{\rm p}=0.35$ and $f_{\rm n}=0.75$, we obtain an evolution of $\sigma_{est}$ for every generated AR that is comparable with the one observed for AR 10268 (see \fig{sigmas}b and \fig{modsig}b). Therefore, using these values for the AR generator parameters, we are introducing perturbations which are comparable with the ones we observed for AR 10268.

%%%%FIGURE modsigmas %%%%%%
%%%%FIGURE modsigmas %%%%%%

\subsection{Inference for generated ARs}
\label{sect_Inference}

 % {\S}{\bf --- One case} \\
We first compare the marginal posteriors obtained with the seed model and one generated AR (blue and green boxes, respectively, in \fig{modbox}).  
The perturbations incorporated with the 4-step generator can modify significantly the deduced AR parameters with respect to the input.  As expected from the results found from the seed model, the changes are dominantly present during the earlier emergence phase for $a$, $B_0$, $d_0$, $R$, $\Nt$, and $\phi$. The $\Nt$ estimation is also affected at the end of the emergence.
This implies that the information lost during the non-linear transformations of the magnetograms cannot be recovered using our inference method.  Still we find the converge of the parameters, except $\Nt$ and $\Bo$,  to the model input during the last part of the emergence.
 
 % {\S}{\bf --- 100 cubes} \\
The analysis of a single case is not representative of all the possible perturbations and their effects on the inferred posteriors. Indeed, random selections of points produce different patterns for the field distributions. By trying different random selections  in the first step of our AR generator, we can statistically explore how the introduced perturbations affect the inference at different stages of the emergence. 
For each magnetogram, we sample the posteriors for 100 generated ARs. 
This means that we apply 100 different patterns that may affect the inferred parameters in various ways. The only condition we impose for the generated ARs, is that $\sigma$ must reach values of at least $200$ G for the negative polarities and $300$ G for the positive ones at some stage of the emergence to guarantee that they differ enough from the seed model.
The orange boxes in \fig{modbox} are constructed as  concatenated distributions of these 100 posteriors, therefore each of these boxes and whiskers contains 200000 samples (2000 for each generated AR). 

 % {\S}{\bf --- 100 cases. General result} \\
Most of the distributions of the 100 cases present a Gaussian-like profile (not shown) implying that the generated perturbations have a random effect on the 8-dimensional posteriors found for each magnetogram. %P To show the effect of these 
We show the means of the 100 AR distributions in \fig{modbox} with orange curves.  The comparison with the original values of the parameters of the seed model (red curves) indicates
systematic deviations. 

%  {\S}{\bf --- how parameters behave at different times} \\
Looking at the distributions at different stages, parameters $a$ and $\Bo$ are well estimated only during the last half of the emergence. During the first half {the radius $a$ is slightly overestimated while $\Bo$ is well underestimated, therefore the FR axial flux (proportional to} $\pi\Bo a^2$ for the half-torus model) is also underestimated during this phase. 
The radius $R$ is well estimated but with a larger uncertainty earlier in the emergence. $\dnot$ is nearly following the linear evolution present in the seed model, but with a relatively large dispersion.
$\Nt$ has the largest deviation from the mean, also with a strong {standard deviation.} Its estimated values are better determined in the central phase of the emergence, when the magnetic tongues are extended and the axial field is strong. 
The dispersion of $\Nt$  
has a great impact on the estimation of $R$ at the first half of the emergence (there, the magnetic tongues modify the bipole size observed in the magnetogram). 

 % {\S}{\bf --- tilt angle} \\
The estimated tilt angle, $\phi$, departs from the input value only at the first magnetogram with largest departure being around $6.5^{o}$.
This initial departure is also observed on other parameters and is due to the largest impact of the perturbations when the AR mean field is weak and comparable to the mean perturbation in terms of $\sigma$.
Later on, the mean $\phi$ of the 100 AR distributions agrees with the input parameter of the seed model reducing also the broadness of the distributions along with the AR emergence.
The obtained values during the full emergence are significantly better than those found from the direct measure of the magnetic barycenters (see white boxes in \fig{modbox}e).  This implies that the Bayesian inference method provides a clear correction of the effect of the tongues on the tilt during the emergence phase. We also include the comparison with CoFFE  in \fig{modbox}e. Though at the beginning of the emergence and up to its first half, a departure from the input values is evident, the results found with CoFFE are much closer than those determined with the barycenter method and clearly converge to the input values {during the second half of the}
emergence phase. 
Although CoFFE achieves a partial correction of the tilt, 
these results show the strength of CoFFE compared to less sophisticated methods, as concluded by \citet{Poisson20a}.

 % {\S}{\bf --- FR center} \\
Finally, the central position parameters, $\xc$ and $\yc$ are well constrained and estimated in all the analyzed cases. As expected, these two parameters are not correlated with the rest of the FR intrinsic parameters.

%%%%%%%%%%%%%%%%%%%%%%%%%%%%%%%%%%%%%%%%%%%%%%%%%%%%%%%%%%
\section{Discussion and Conclusions}   \label{sect_Conclusion}

%  {\S}{\bf --- In this work -- summary} \\
In this work we test a new method based on a Bayesian statistical approach to derive active region global parameters from LOS magnetograms. 
We use a synthetic half-torus model successfully employed in previous works to interpret the magnetic tongues and their evolution during the emergence of bipolar ARs \citep{Luoni11,Poisson15a}.  This model depends on eight free parameters.  Their most probable values and statistical distributions (called posteriors) are derived applying this new Bayesian approach to observed magnetograms. The magnetograms are treated independently during the AR emergence, so the effect of the statistical procedure on the determination of the eight-model parameters can be tested at different stages of the analyzed evolution.

 % {\S}{\bf --- magnetogram model} \\
Despite the simplicity of the half-torus model, we are able to determine the large-scale properties of AR 10268 from its LOS magnetic field. More precisely, we reproduce its global characteristics, such as the elongation of the polarities and the inclination of the PIL on the bipole axis. Moreover, the evolution of both observed and modeled magnetograms are comparable at any stage of the AR emergence (see movie mov-fig3.mp4 from the supplementary material). Using this model we are also able to reduce the contribution of the background flux, not belonging to the AR, finding a more balanced evolution of the AR main positive and the negative flux. We establish an upper bound for the observed standard deviation ($\sigma$) between the model and the observations that ranges between $100$ G and $300$ G. This large {standard deviation} is partly due to the asymmetry that exists between the leading and following polarities, both in their extension and small scale field perturbations.

%  {\S}{\bf --- AR generator} \\
To determine the effect of the small scale perturbations on the posterior we create an AR generator to mimic realistic field observations starting from a smooth seed model.  The aim is to apply strong modifications at small scales while keeping the global properties {almost unchanged.} %intact. 
For the generated ARs we use the same $\sigma$ value obtained for AR 10268 to maintain the modifications introduced to the magnetic field at a reasonable level. The inference is made over 100 generated AR emergences using the same method applied to AR 10268 magnetograms. This test shows that the introduced perturbations could be large enough to shift the posterior distributions of the parameters for one simulated AR by more than one inter-quartile range from the input seed model parameters.

%  {\S}{\bf --- Comparison between model and AR 10268 results} \\
The marginal posteriors {are defined for each model parameter by integrating the full posterior over the remaining 7 parameters. Such marginal posteriors,} shown in \fig{boxplot} as boxplots, display a large variation (both in {their mean and standard deviation}) from the parameters of the model during the AR 10268 emergence. Despite these temporal variations {are not included within the half-torus model,} 
most of them are compatible with small scale random perturbations of the field distribution, as seen for the generated AR marginal posteriors of \fig{modbox}. 

Comparing the results derived from the analysis using the AR generator to those coming directly from the application of the Bayesian inference method to AR 10268, we find a good correspondence that can we summarized as:
\begin{itemize}
    \item The FR cross-section radius, \ie\ the parameter $a$ in \fig{boxplot}a, has a strong variation and large dispersion during the first third of the emergence and then it converges to a constant value for the rest of its evolution. The first variation is characterized by a fast expansion reaching values above the later convergence value. There is also an overestimation, though milder, of $a$ during the first half of the emergence for the generated AR selected (green line, \fig{modbox}a). This effect is weaker when a set of 100 ARs is selected (orange line).  Later on, $a$ uniformly converges to the input value. We conclude that the evolution of $a$ observed for AR 10268 has a contribution induced by {the sensitivity of the half-torus model to} small scale perturbations.
    
    \item  The first small values of the axial field strength, $\Bo$ in \fig{boxplot}b, could, at least partly,  be a bias of the method, since a similar, although less important, behavior is present on the posterior of the seed model {(blue boxes in \fig{modbox}b).}  Moreover, this bias becomes stronger once the small scale perturbations are introduced to the seed model (orange boxes in \fig{modbox}b). The difficulty to determine both $a$ and $\Bo$ during the first half of the emergence is due to the partial emergence of the FR and the projection of the axial field on the LOS direction. 
    The anti-correlation between $a$ and $\Bo$, obtained for AR 10268 and the generated ARs, suggest that the axial flux can be a less biased parameter than $\Bo$. In other words, despite $a$ and $\Bo$ are not well estimated separately, the combination of $\Bo a^2$ might be a more robust parameter for the FR model.
    
    \item  The evolution of the main FR radius, $R$ in \fig{boxplot}c, can potentially be an artifact of the method, since an increase of the distance between the main polarities can be reproduced with an increase of $R$ at constant $\dnot$. 
    Indeed,  although the width of the distributions of $R$ remains mostly constant during emergence for the 100 generated ARs, the marginal posteriors of the single generated AR example shown in \fig{modbox}c presents similar temporal variation as the one obtained for AR 10268 (green boxes). This implies that the variation of $R$ for AR 10268 is {potentially} within the framework of the torus model considering the effect of the small scale field perturbations. 
    
    \item The evolution of the bipole tilt, $\phi$ in \fig{boxplot}e, shows a clear clockwise rotation of AR 10268, which could well be associated to the torsion (or writhe) of the FR axis.  The estimation obtained is also close to the results obtained with CoFFE \citep{Poisson20a} and significantly different from the tilt obtained by computing the magnetic barycenters, which are affected by the magnetic tongues. \fig{modbox}e shows the improvement due to the removal of the effect of the tongues on the estimation of the tilt. Moreover, the Bayesian approach proves to be more stable and with a more robust estimation than CoFFE at the first half of the emergence. {We conclude that the clockwise rotation of AR 10268 is not due to the evolution of the magnetic tongues but the actual measure of the full AR rotation}.
    
    \item The evolution of the twist number, $\Nt$ in \fig{boxplot}f, has a similar behavior to the simulated ARs in \fig{modbox}f, with lower values near the beginning and the end of the emergence. For the generated ARs the best estimations of $\Nt$ are obtained around the middle of the emergence phase when the elongation of the polarities is maximum. The observed variation is consistent with the value of the twist number computed for AR 10268 by \citet{Poisson15a}.  

   \item Finally the estimation of the AR center, corresponding to parameters $x_c$ and $y_c$, provides the most robust result for both AR 10268 and the generated ARs (\fig{boxplot}g,h and \fig{modbox}g,h respectively). This is expected because these parameters are not intrinsic FR quantities, therefore their correlations with the other model parameters is negligible. 
\end{itemize}

The PYMC3 software, used in this work, is shown to be efficient and versatile to be adapted to the problem needs (high model dimension) with diverse tools for tensor computation, sampling and Bayesian inference. Other samplers, e.g. Hamiltonian Monte Carlo algorithms, and different prior constructions including parameter correlations \citep[LKG distributions;][]{LKG1989}, also available within the software package, are worth to be tested in future works.
On the other hand, the present Bayesian analysis implementation allows us to explore the half-torus model, obtaining a first description of the parameters and their correlations. Moreover, this method also quantifies the statistical distributions of the output parameters, allowing us to better characterize its limitations, which are mainly present during the early emergence phase. 

The direct comparison of a model with observations introduces a new method to derive the temporal evolution of the global parameters of an emerging AR to a precision level surpassing previous methods. In particular, the tilt angle obtained from the model for AR 10268 in the early emergence phase corrects substantially the effect of the magnetic tongues. 
{The intrinsic tilt of ARs is an important observational constraint for flux-transport dynamo models, since the AR tilt is linked, via diffusion and meridional circulation, to the formation and evolution of the polar field during solar cycles \citep[e.g.][]{Wang17}. Then, the application of the Bayesian method presented here to a large set of ARs, in the spirit of the work of \citet{Poisson20a}, is expected to improve our knowledge of the tilt angle dependence with latitude.} 

% rotation of ARs
{As a FR emerges, the magnetic tongues evolve producing an apparent rotation of the magnetic bipole.  By removing this effect, the Bayesian method provides a better estimation of the actual intrinsic rotation of the bipole. This rotation is around $20\degree$ for AR 10268, a factor two smaller than the tilt evolution observed with the effect of the magnetic tongues included. The amount of rotation and its direction { has a broad distribution in ARs \citep{Lopez-Fuentes03}, but its physical origin is still a matter of debate}. For example, the tilt angle evolution obtained for AR 10268 (clockwise rotation) suggests a deformation of the FR axis { consistent with a negative writhe,} but the origin of this writhing remains unexplained. We expect that the application of {the just developed and tested} method to a larger set of bipolar ARs will help to understand the physical origin of the rotation of AR bipoles.}

\textbf{Acknowledgments}\\
CHM, MLF and MP acknowledge grants PICT-2020-SERIEA-03214 (ANPCyT), PIP 11220200100985 (CONICET) and UBACyT 20020170100611BA. CHM, FG, MLF and MP are members of the Carrera del Investigador Cient\'\i fico of the Consejo Nacional de Investigaciones Cient\'\i ficas y T\'ecnicas (CONICET). 
We thank MDI team for providing the magnetograms used in this study. 
We recognize the collaborative and open nature of knowledge creation and dissemination under the control of the academic community, as expressed by Camille No\^{u}s at http://www.cogitamus.fr/indexen.html.

%% Bibliography
%% Author year style
\bibliographystyle{aa}
%\biboptions{authoryear}
\bibliography{biblio3}

\begin{appendix}

\section{Half-torus model} \label{app_torus}

%%%%FIGURE models boxplot %%%%%%
\begin{figure*}[!t]
\centering
\includegraphics[width=0.9\textwidth]{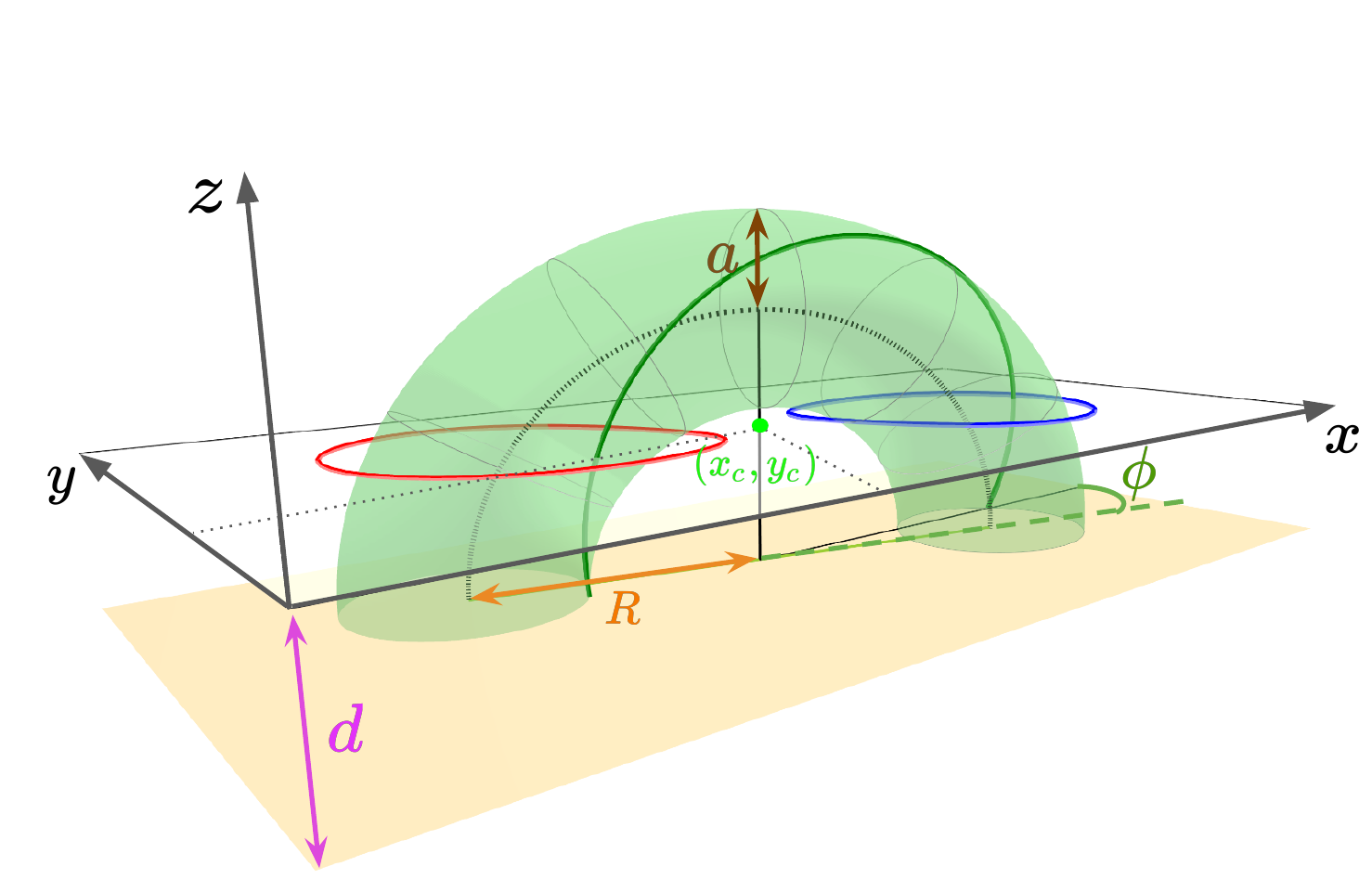}
\caption{
Diagram of the field morphology of the half-torus model described in \sect{torus}. This diagram corresponds to the test model presented in  \sect{simplecase}. The dotted line corresponds to the torus axis with a field strength $\Bo = 3000$ G. The light green surface shows the tube isosurface defined by a small radius $a = 4.5$ and large radius $R = 12$, both in pixel units.  The $x_c$ and $y_c$ parameters correspond to the torus central position in pixel units in the $xy-$plane. $\phi$ is the tube tilt angle with respect to $x-$axis. The photospheric plane is located at a height $d=6.5$ with respect to the torus center.
The red and blue contours over this plane indicates the projection of the torus field in the $z-$direction with a field strength of $500$ G and $-500$ G, respectively. 
}
\label{fig_FR_sketch}
\end{figure*}
%%%%FIGURE models boxplot %%%%%%

{We summarize below the flux rope (FR) model used in this paper and developed by \citet{Luoni11}.  The  model was designed to be the simplest possible while being able to describe the key features of emerging ARs, in particular magnetic tongues.  A visualization of the model and its parameters is shown in Figure~\ref{fig_FR_sketch}.
The magnetic field has a toroidal geometry with an invariance set along the FR axis.  The magnetic twist is set uniform both along and across the FR axis.  The axial 
field $B_{\alpha}$ and the azimuthal field $B_{\theta}$ are defined as 

  \BA \label{Eq-Bphi}
     B_{\alpha} &=& B_{0} \exp (-(\rho/a)^2)  \,,  \\
     B_{\theta} &=& \frac{2 \rho \Nt}{R + \rho \cos (\theta)} B_{0} \exp (-(\rho/a)^2)  \,, 
   \EA
   
\noindent where $\{\rho,\alpha,\theta\}$ are the coordinates for the torus, in which $\rho$ is the distance to the torus axis, $\alpha$ is the location along the torus axis, and $\theta$ corresponds to the rotation angle around the torus axis. $a$ is the typical small radius of the flux rope and $B_{0}$ is the field strength on the axis.  The FR axis has a radius $R$, and the center of the torus is located at a depth $d$ below the plane where we compute the magnetic field to simulate magnetograms.  The number of field line turns (twist) along the half-torus is noted as $\Nt$.   Next, the horizontal position of the torus center is defined with the parameters $\xc$ and $\yc$ and the orientation of the FR with the tilt $\phi$. Finally, the model is fully defined by 8 parameters.
}

\end{appendix}

\end{document}

%% file: mod_AR_definitions.tex
\renewcommand{\mp}[1]{{\color{blue}{#1}}}
\newcommand{\degree}{^\circ} 
\newcommand{\BE}{\begin{equation}}
\newcommand{\EE}{\end{equation}}
\newcommand{\BA}{\begin{eqnarray}}
\newcommand{\EA}{\end{eqnarray}}

\newcommand{\fig}[1]{Fig.~\ref{fig_#1}}

\newcommand{\sect}[1]{Sect.~\ref{sect_#1}}
\newcommand{\app}[1]{Appendix~\ref{app_#1}}

\newcommand{\eq}[1]{Eq.~(\ref{eq_#1})}

\newcommand{\eg}{{\it e.g.},}

\newcommand{\ie}{{\it i.e.}}

\newcommand{\Bo}{B_{\rm 0}}
\newcommand{\dnot}{d_{\rm 0}}
\newcommand{\Nt}{N_{\rm t}}
\newcommand{\yc}{y_{\rm c}}
\newcommand{\xc}{x_{\rm c}}

\newcommand{\Bz}{B_{\rm z}}

\newcommand{\comment}[1]{}

\newcommand{\Bj}[1]{B_{#1,j}}
\newcommand{\cL}{\mathcal{L}}

\newcommand{\vp}{\vec{\mathfrak{p}}}

\newcommand{\Mpi}{M_{\vp_i}}
\newcommand{\Mo}{M_o}

\newcommand\maps{\ref@jnl{M\&PS}}% Meteoritics and Planetary Science
\newcommand\aas{\ref@jnl{AAS Meeting Abstracts}}% American Astronomical Society Meeting Abstracts
\newcommand\dps{\ref@jnl{AAS/DPS Meeting Abstracts}}% American Astronomical Society/Division for Planetary Sciences Meeting Abstracts